\documentclass[11pt,a4paper]{article}
\usepackage[utf8]{inputenc}
\usepackage[T1]{fontenc}
\usepackage[greek,english]{babel}

\usepackage[hyperref,acceptedWithA]{tacl2018v2}
\hypersetup{hypertexnames=false}
\DeclareMathAlphabet{\mathgtt}{LGR}{cmtt}{m}{n}

\usepackage{comment}
\usepackage{times}
\usepackage{latexsym}

\usepackage{xcolor}
\definecolor{darkgreen}{rgb}{0.0,0.5,0.0}
\definecolor{darkblue}{rgb}{0.0,0.0,0.5}
\definecolor{darkred}{rgb}{0.55, 0.0, 0.0}
\definecolor{cornsilk}{rgb}{1.0, 0.97, 0.86}
\definecolor{desertsand}{rgb}{0.93, 0.79, 0.69}

\usepackage{amsfonts}
\usepackage{amsmath}
\usepackage{amssymb}
\usepackage{amsthm}

\usepackage{mathtools}
\usepackage{stmaryrd}
\usepackage{xspace}
\usepackage{booktabs}
\usepackage{relsize}
\usepackage[normalem]{ulem}

\usepackage{inconsolata}

\usepackage{listings}
\usepackage{xcolor}
\usepackage[LGR,T1]{fontenc}
\usepackage{amsmath}
\usepackage{xcolor}

\DeclareMathAlphabet{\mathgtt}{LGR}{cmtt}{m}{n}

\usepackage{listings}
\usepackage{xcolor}

\definecolor{darkgreen}{rgb}{0.0,0.5,0.0}
\definecolor{darkblue}{rgb}{0.0,0.0,0.5}
\definecolor{aggblue}{rgb}{0,0,.65}
\definecolor{vargreen}{rgb}{0,.50,.18}

\newcommand{\var}[1]{\texttt{\dinline{#1}}\xspace}

\newcommand{\vI}{\var{I}}
\newcommand{\vJ}{\var{J}}
\newcommand{\vK}{\var{K}}

\newcommand{\vN}{\var{N}}

\newcommand{\vW}{\var{W}}
\newcommand{\vX}{\var{X}}
\newcommand{\vY}{\var{Y}}
\newcommand{\vZ}{\var{Z}}

\newcommand{\dd}[1]{{\small\tt #1}}
\newcommand{\dinline}[1]{\text{\lstinline[language=dyna]{#1}}}

\lstdefinelanguage{dyna}{
  sensitive=true,
  morecomment=[l]{\%},
  morestring=[b]",
  classoffset=0,
  morekeywords={for,with\_key,arg,in,out,is},
  keywordstyle=\bf\tt\color{aggblue},
  classoffset=2,
  morekeywords={A,B,C,D,E,F,G,H,I,J,K,L,M,N,O,P,Q,R,S,T,U,V,W,X,Y,Z,Xs,Ys,Zs,DX,DY,DZ,\_},
  keywordstyle=\color{vargreen},
  classoffset=0,
  keywordstyle=\color{blue},
  literate=
    {alpha}{$\mathgtt{a}$}2
    {beta}{{{$\mathgtt{b}$}}}2
    {gamma}{{{$\mathgtt{g}$}}}2
    {epsilon}{{{$\mathgtt{e}$}}}2
    {.}{{{\color{aggblue}.}}}2
    {input:}{{{\color{aggblue}input:}}}2
    {params:}{{{\color{aggblue}params:}}}2
    {output:}{{{\color{aggblue}output:}}}2
    {:-}{{{\color{aggblue}:-}}}2
    {?}{{{\color{aggblue}?}}}2
    {<}{{{\color{aggblue}<}}}2
    {>}{{{\color{aggblue}>}}}5
    {=}{{{\color{aggblue}=}}}2
    {min=}{{{\color{aggblue}min=}}}2
    {max=}{{{\color{aggblue}max=}}}2
    {~>}{{{$\rightarrow$}}}5
    {<~}{{{$\leftarrow$}}}5
    {+=}{{{\color{aggblue}+=}}}2
    {?=}{{{\color{aggblue}?=}}}2
    {*}{{{\color{aggblue}*}}}1
    {-}{{{\color{aggblue}-}}}4
    {+}{{{\color{aggblue}+}}}1
    {^}{{{\color{aggblue}\textasciicircum}}}2
    {_0}{{$_0$}}2
    {_1}{{$_1$}}2
    {_2}{{$_2$}}2
    {_3}{{$_3$}}2
    {_4}{{$_4$}}2
    {_5}{{$_5$}}2
    {_6}{{$_6$}}2
    {_7}{{$_7$}}2
    {_8}{{$_8$}}2
    {_9}{{$_9$}}2
    {_i}{{$_i$}}2
    {_j}{{$_j$}}2
    {_k}{{$_k$}}2
    {_K}{{$_K$}}2
    {_n}{{$_n$}}2
    {_N}{{$_N$}}2
    {ldots}{{$\myDots$}}2
}

\lstnewenvironment{dynaex}[2][]
                  {%
                    \lstset{
                      language={dyna},
                      classoffset=2,
                      morekeywords={#2},
                      keywordstyle=\color{vargreen},
                      #1
                    }
                    \vspace{-2pt}
                  }
                  {%
                    \vspace{-2pt}
                  }

\usepackage{listings}
\usepackage{xcolor}
\usepackage[LGR,T1]{fontenc}
\usepackage{amsmath}
\usepackage{xcolor}
\definecolor{darkgreen}{rgb}{0.0,0.5,0.0}
\definecolor{darkblue}{rgb}{0.0,0.0,0.5}
\definecolor{darkred}{rgb}{0.55, 0.0, 0.0}
\definecolor{cornsilk}{rgb}{1.0, 0.97, 0.86}
\definecolor{desertsand}{rgb}{0.93, 0.79, 0.69}
\definecolor{darkcyan}{rgb}{0.0, 0.55, 0.55}
\definecolor{darkkhaki}{rgb}{0.74, 0.72, 0.42}
\definecolor{darkchestnut}{rgb}{0.6, 0.41, 0.38}
\definecolor{charcoal}{rgb}{0.21, 0.27, 0.31}
\definecolor{chromeyellow}{rgb}{1.0, 0.65, 0.0}
\definecolor{cadmiumgreen}{rgb}{0.0, 0.42, 0.24}
\definecolor{camouflagegreen}{rgb}{0.47, 0.53, 0.42}
\definecolor{burntorange}{rgb}{0.8, 0.33, 0.0}
\definecolor{chocolate}{rgb}{0.82, 0.41, 0.12}
\definecolor{phthalogreen}{rgb}{0.07, 0.21, 0.14}
\definecolor{darksienna}{rgb}{0.24, 0.08, 0.08}
\definecolor{darkraspberry}{rgb}{0.53, 0.15, 0.34}

\DeclareMathAlphabet{\mathgtt}{LGR}{cmtt}{m}{n}

\newcommand{\tinline}[1]{\text{\lstinline[language=type]{#1}}}

\newcommand{\est}[1]{\ensuremath{\overline{\tt #1}}}

\newcommand{\paramColor}[0]{\color{darkraspberry}}

\lstdefinelanguage{type}{
  sensitive=true,
  morecomment=[s]{`}{`},
  morecomment=[l]{\%},
  morestring=[b]",
  classoffset=0,
  morekeywords={for,with\_key,arg,in,is,out},
  keywordstyle=\bf\tt\color{aggblue},
  classoffset=2,
  morekeywords={A,B,C,D,E,F,G,H,I,J,K,L,M,N,O,P,Q,R,S,T,U,V,W,X,Y,Z,Xs,Ys,Zs,DX,DY,DZ,\_},
  keywordstyle=\color{vargreen},
  classoffset=0,                          
  keywordstyle=\paramColor,
  morekeywords={a,b,c,k,w,n,ks,int,str,item,times,lessthan,plus},
  classoffset=3,                          
  keywordstyle=\est,
  morekeywords={score,shift,edge,start,stop,reduce,word,need,goal,len,b,h},
  classoffset=4,  
  keywordstyle=\color{blue},
  basicstyle={\ttfamily\footnotesize},
  xleftmargin={0.5cm},
  numbers=left,
  framerule=0pt
  frame=none,
  stepnumber=1,
  firstnumber=1,
  numberfirstline=true,
  tabsize=2,
  showtabs=false,
  showspaces=false,
  showstringspaces=false,
  extendedchars=true,
  breaklines=true,
  columns=fullflexible,
  keepspaces=true,
  escapeinside={@}{@},
  firstnumber=last,
  commentstyle=\color{black!50},
  numberstyle=\tiny\color{black!50},
  literate=
    {np}{\dd{np}}2
    {alpha}{\est{\mathgtt{a}}}2
    {beta}{\est{\mathgtt{b}}}2
    {gamma}{\est{\mathgtt{g}}}2
    {shift}{\est{shift}}2
    {lessthan}{{\color{darkraspberry}lessthan}}2
    {epsilon}{\mathgtt{e}}2
    {true}{{\color{aggblue}true}}2
    {false}{{\color{aggblue}false}}2
    {.}{{{\color{aggblue}.}}}2
    {params:}{{{\color{aggblue}params:}}}2
    {:-}{{{\color{aggblue}:-}}}2
    {?}{{{\color{red}?}}}2
    {<}{{{\color{aggblue}<}}}5
    {>}{{{\color{aggblue}>}}}5
    {<=}{{{\color{aggblue}\ensuremath{\le}}}}5
    {>=}{{{\color{aggblue}\ensuremath{\ge}}}}5
    {<==}{{{\color{aggblue}<==}}}2
    {=}{{{\color{aggblue}=}}}2
    {min=}{{{\color{red}min=}}}2
    {max=}{{{\color{red}max=}}}2
    {~>}{{{\ensuremath{\rightarrow}}}}5
    {<~}{{{\ensuremath{\leftarrow}}}}5
    {+=}{{{\color{red}+=}}}2
    {?=}{{{\color{red}?=}}}2
    {*}{{{\color{red}*}}}1
    {-}{{{\color{red}-}}}4
    {+}{{{\color{vargreen}+}}}1
    {^}{{{\color{aggblue}\textasciicircum}}}2
    {_0}{\ensuremath{_0}}2
    {_1}{\ensuremath{_1}}2
    {_2}{\ensuremath{_2}}2
    {_3}{\ensuremath{_3}}2
    {_4}{\ensuremath{_4}}2
    {_5}{\ensuremath{_5}}2
    {_6}{\ensuremath{_6}}2
    {_7}{\ensuremath{_7}}2
    {_8}{\ensuremath{_8}}2
    {_9}{\ensuremath{_9}}2
    {_i}{\ensuremath{_i}}2
    {_j}{\ensuremath{_j}}2
    {_k}{\ensuremath{_k}}2
    {_m}{\ensuremath{_m}}2
    {_K}{\ensuremath{_K}}2
    {_N}{\ensuremath{_N}}2
    {_M}{\ensuremath{_M}}2
    {ldots}{{$\myDots$}}2
}

\lstnewenvironment{typeex}[2][]
                  {%
                    \lstset{
                      language={type},
                      morekeywords={for,out,with\_key,arg},
                      classoffset=6,
                      morekeywords={#2},
                      keywordstyle=\color{vargreen},
                      #1
                    }
                    \vspace{-2pt}
                  }
                  {%
                    \vspace{-2pt}
                  }

\newcommand{\sym}[1]{{\color{burntorange}\ensuremath{#1}}}

\newcommand{\Sg}[0]{\sym{g}}

\newcommand{\Sk}[0]{\sym{k}}

\newcommand{\Sn}[0]{\sym{n}}

\newcommand{\Sw}[0]{\sym{w}}

\usepackage{tikz}
\usetikzlibrary{tikzmark}
\usetikzlibrary{calc}

\newcommand{\pgftextcircled}[1]{
    \setbox0=\hbox{#1}%
    \dimen0\wd0%
    \divide\dimen0 by 2%
    \begin{tikzpicture}[baseline=(a.base)]%
        \useasboundingbox (-\the\dimen0,0pt) rectangle (\the\dimen0,1pt);
        \node[circle,draw,outer sep=0pt,inner sep=0.1ex] (a) {#1};
    \end{tikzpicture}
}

\newcommand{\nonsemizero}[0]{\ensuremath{\mbox{\text{non-}\!\semizero}}\xspace}
\newcommand{\semizero}[0]{\,{\smaller\pgftextcircled{\smaller $0$}}\,}
\newcommand{\semione}[0]{\,{\smaller\pgftextcircled{\smaller $1$}}\,}

\newcommand{\codefont}{\fontfamily{lmtt}\selectfont}
\newcommand{\data}[1]{\texttt{\codefont#1}}
\newcommand{\code}[1]{\data{#1}}
\newcommand{\mvar}[1]{\code{\textit{#1}\hspace{-1pt}}}

\newcommand{\dhead}[0]{\mvar{h}}
\newcommand{\db}[0]{\mvar{b}}

\newcommand{\ddh}[0]{\mvar{h}}
\newcommand{\ddc}[0]{\mvar{c}}
\newcommand{\ddb}[0]{\mvar{b}}

\newcommand{\cutforspace}[1]{}

\usepackage[disable]{todonotes}
\makeatletter
\newcommand*\iftodonotes{\if@todonotes@disabled\expandafter\@secondoftwo\else\expandafter\@firstoftwo\fi}
\newcommand{\noindentaftertodo}{\iftodonotes{\noindent}{}}
\newcommand{\note}[4][]{\todo[author=#2,color=#3,size=\scriptsize,caption={},#1]{#4}}

\newcommand{\response}[1]{\vspace{3pt}\hrule\vspace{3pt}\textbf{#1:}}

\newcommand{\timv}[2][]{\note[#1]{timv}{magenta!40}{#2}}
\newcommand{\Timv}[2][]{\timv[inline,#1]{#2}\noindentaftertodo}

\newcommand{\jason}[2][]{\note[#1]{jason}{green!40}{#2}}
\newcommand{\Jason}[2][]{\jason[inline,#1]{#2}\noindentaftertodo}

\usepackage{algorithm}
\usepackage[noend]{algpseudocode}
\algrenewcommand\algorithmicindent{1.0em}%

\newcommand{\rightcomment}[1]{{\color{gray} \(\triangleright\) {\smaller\textit{#1}}}}
\algrenewcommand{\algorithmiccomment}[1]{\hfill \rightcomment{#1}}
\algnewcommand{\LineComment}[1]{\State\rightcomment{#1}}
\algnewcommand{\LinesComment}[1]{\State\rightcomment{\parbox[t]{\linewidth-\leftmargin-\widthof{\(\triangleright\) 111}}{#1}}}
\algrenewcommand\alglinenumber[1]{{\tiny\color{black!50}#1.}\hspace{-2pt}}

\newcommand{\algorithmicfunc}[1]{\textbf{def} {#1}:}
\algdef{SE}[FUNC]{Func}{EndFunc}[1]{\algorithmicfunc{#1}}{}
\makeatletter
\ifthenelse{\equal{\ALG@noend}{t}}%
  {\algtext*{EndFunc}}
  {}%
\makeatother

\newcommand{\notation}[1]{{\color{darkblue}#1}}
\newcommand{\mysf}[1]{\notation{\textsf{\smaller #1}}}

\newcommand{\dtimes}{{\color{darkblue}\texttt{*}}}
\newcommand{\dplus}{{\color{darkblue}\texttt{+}}}

\newcommand{\dopluseq}{{\color{darkblue}\oplus\texttt{=}}}
\newcommand{\doplus}{{\color{darkblue}\oplus}}
\newcommand{\dotimes}{{\color{darkblue}\otimes}}

\newcommand{\CC}[0]{\mathcal{C}}

\newcommand{\head}[1]{\mysf{head}{(#1)}}
\newcommand{\body}[1]{\mysf{body}{(#1)}}
\newcommand{\vars}[1]{\mysf{vars}{(#1)}}

\newcommand{\Prog}[0]{\mathcal{P}}
\newcommand{\Data}[0]{\mathcal{D}\xspace}
\newcommand{\BProg}[0]{\overline{\Prog}}
\newcommand{\Shape}[0]{\mathcal{S}\xspace}

\newcommand{\Subst}[0]{\mysf{subst}}
\newcommand{\subst}[2]{\Subst(#1, #2)}

\newcommand{\Fresh}[0]{\mysf{fresh}}
\newcommand{\fresh}[1]{\Fresh{(#1)}}

\newcommand{\Unify}[0]{\mysf{unify}}
\newcommand{\unify}[1]{\Unify{(#1)}}

\newcommand{\colondash}[0]{\xspace\texttt{:-}\xspace}

\newcommand{\theHerb}[0]{\mathbb{H}}
\newcommand{\Values}[0]{\mathbb{V}}

\newcommand{\support}[1]{\mysf{nz}(#1)}

\newcommand{\myfunc}[1]{\notation{\textsf{\smaller #1}}}
\newcommand{\LookupT}[0]{\myfunc{lookup}}
\newcommand{\FixedPointT}[0]{\myfunc{step\ensuremath{^*}}}

\newcommand{\Inflate}[0]{\myfunc{step}}
\newcommand{\Expand}[0]{\myfunc{step\_rule}}

\newcommand{\Propagate}[0]{\myfunc{propagate}}
\newcommand{\Relax}[0]{\myfunc{relax}}

\newcommand{\Bigo}[0]{\mathcal{O}}
\newcommand{\bigo}[1]{\Bigo\!\left(#1\right)}

\usepackage{microtype}
\usepackage[inline]{enumitem}

\newcommand{\vtheta}{\boldsymbol{\theta}}

\newcommand{\sem}[1]{\left\llbracket #1 \right\rrbracket}

\newcommand{\tuple}[1]{\langle #1 \rangle}
\newcommand{\Set}[1]{\left\{ #1 \right\}}

\newcommand{\myDots}{\ifmmode\mathinner{\ldotp\kern-0.2em\ldotp\kern-0.2em\ldotp}\else.\kern-0.13em.\kern-0.13em.\fi}

\newcommand{\defeq}[0]{\mathrel{\stackrel{\textnormal{\tiny def}}{=}}}

\usepackage{thmtools}

\newtheorem{myexample}{Example}

\newtheorem{defin}{Definition}
\theoremstyle{definition}

\usepackage{cleveref}   
\Crefname{ALC@unique}{Line}{Lines}
\crefname{section}{\S}{\S}
\Crefname{section}{\S}{\S}
\crefformat{section}{#2\S#1#3}
\crefrangeformat{section}{\S#3#1#4--#5#2#6}
\Crefrangeformat{section}{\S#3#1#4--#5#2#6}
\crefmultiformat{section}{\S#2#1#3}{ and~\S#2#1#3}{, \S#2#1#3}{ and~\S#2#1#3}
\Crefmultiformat{section}{\S#2#1#3}{ and~\S#2#1#3}{, \S#2#1#3}{ and~\S#2#1#3}
\crefrangemultiformat{section}{\S\S#3#1#4--#5#2#6}{ and~\S\S#3#1#4--#5#2#6}{, \S\S#3#1#4--#5#2#6}{ and~\S\S#3#1#4--#5#2#6}
\Crefrangemultiformat{section}{\S\S#3#1#4--#5#2#6}{ and~\S\S#3#1#4--#5#2#6}{, \S\S#3#1#4--#5#2#6}{ and~\S\S#3#1#4--#5#2#6}
\crefrangeformat{line}{lines #3#1#4--#5#2#6}
\crefname{table}{Table}{Tables}
\crefname{figure}{Fig.}{Fig.}
\crefname{algorithm}{Alg}{Alg}
\crefname{algorithm}{Alg}{Alg}
\crefname{line}{line}{lines}
\crefname{appendix}{\S\!\!}{\S\!\!}
\crefname{thm}{Theorem}{}
\crefname{prop}{Prop.\@}{Props.\@}
\crefname{defin}{Definition}{Definitions}
\crefname{lemma}{Lemma}{Lemmata}
\crefname{cor}{Corollary}{Corollaries}
\crefname{equation}{}{}
\crefname{myexample}{Example}{Examples}

\newcommand{\defn}[1]{\textbf{#1}}
\everypar{\looseness=-1}

\title{Automating the Analysis of Parsing Algorithms \\ (and other Dynamic Programs)}

\author{ Tim Vieira \\
   Johns Hopkins University \\
  \href{mailto:tim.f.vieira@gmail.com}{\tt tim.f.vieira@gmail.com} \\
  \And
  Ryan Cotterell \\
  ETH Z\"{u}rich \\
  \href{mailto:ryan.cotterell@inf.ethz.ch}{\tt \ \ ryan.cotterell@inf.ethz.ch}  \\
  \And
  Jason Eisner \\
  Johns Hopkins University \\
  \href{mailto:jason@cs.jhu.edu}{\tt jason@cs.jhu.edu}
}

\date{}

\newcommand{\citestamp}{
  \AddToShipoutPicture*{%
    \setlength{\unitlength}{1mm}
    \put(105,14){\makebox(0,0){\parbox{\textwidth}{\footnotesize\color{darkblue}\textbf{Authors' Note:} This manuscript
        may be cited as Vieira et al.\ (2021).  We submitted it in November 2021 to the journal
        \emph{Transactions of the Association for Computational Linguistics},
        which conditionally accepted it 2 months later.  However, we never made the required or suggested expository
        revisions: our apologies to the TACL editors and reviewers.  An expanded version can be found as chapter 6 of the first author's Ph.D. dissertation \cite{vieira-2023}.}}}
  }
}

\begin{document}
\maketitle

\begin{abstract}
Much algorithmic research in NLP aims to efficiently manipulate rich formal structures.
An algorithm designer typically seeks to provide guarantees about their proposed algorithm---for example, that its running time or space complexity is upper-bounded as a certain function of its input size.  
They may also wish to determine the necessary properties of the quantities derived by the algorithm to synthesize efficient data structures and verify type errors.
In this paper, we develop a system for helping programmers to perform these types of analyses.  We apply our system to a number of NLP algorithms and find that it successfully infers types, dead and redundant code, and parametric runtime and space complexity bounds.

\includegraphics[height=1em]{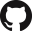} {\tt\fontsize{9.15}{11}\selectfont\href{https://github.com/timvieira/dyna-pi}{https://github.com/timvieira/dyna-pi}}

\end{abstract}

\citestamp

\section{Introduction}\label{sec:intro}
The goal of this work is to further the pursuit of declarative programming for NLP \citep{pereira2002prolog,cgr09compiler}.
The promise of declarative programming is that it enables the programmer to specify their algorithm in a high-level language without worrying about low-level implementation details and instead focus on the correctness of the algorithm itself.
This paper focuses on Dyna \citep{eisner-goldlust-smith-2005}, a declarative programming language that was designed to appeal to NLP researchers.
Dyna is a concise language for implementing dynamic programs, which have long been the workhorse of structured prediction in many NLP problems.
Specifically, NLP makes heavy use of probabilistic models based on rich linguistic formalisms, e.g. 
finite-state transduction \cite{mohri-1997-finite},
context-free parsing \cite{goodman-1999-semiring},
dependency parsing \cite{eisner-1996-three}
and mildly context-sensitive parsing \cite{vijay-shanker-weir-1989-recognition},
in which inference requires dynamic programming.
Such algorithms are often difficult to implement correctly and efficiently.\footnote{These classical NLP algorithms are still often used for deep structured prediction, when the score of a structure decomposes into local terms that are computed by neural networks \citep{durrett-klein-2015-neural,rastogi2016weighting,lee2016global,%
kim2017structured-attention,%
qi2020stanza,rush-2020-torch-struct}.}\looseness=-1

Now we turn to static analysis of these declarative programs---e.g., type analysis, runtime complexity analysis, and space complexity analysis \citep{cousot77abstract,types}.
Such analysis is important for understanding the space of possible program behaviors on unknown input data.\footnote{\label{fn:compiler-optimizations}Although we do not explore it in this paper, we also mention that static program analysis is essential for compiling efficient code.  Type inference, for example, allows the compiler to synthesize efficient data structures for representing internal objects and efficient code for pattern-matching against them.\looseness=-1
}\Jason{add a mention to footnote of compile-time assertion checking---both explicit user assertions, and implicit assertions about aggregators not clashing and the = aggregator having at most one aggregand; can also prune some deadcode paths where the test must fail}
\emph{Automated} program analysis aids programmers because \emph{manual} program analysis may be tedious and error-prone.
In the context of NLP, automated program analysis of dynamic programs (for instance, as specified in Dyna) has the potential to help NLP algorithms researchers correctly and speedily analyze their creations.\cutforspace{Automatic analysis is especially important for large programs, such as those derived automatically by program transformation, such as those in \citet{eisner-blatz-2007}, \citet{cohen09products}, or \citet{vieira2021searching}.  Program analysis is also essential to tidying up these generated programs, e.g., by removing rules that will never fire (dead code elimination).}
Indeed, the NLP literature provides examples
where such automated analysis could have been of use.
For example, \citet{huang-sagae-2010-dynamic} derive a projective dependency parsing algorithm with a stated complexity of $\bigo{n^7}$; subsequent work \cite{shi-etal-2017-fast} improved the analysis to discover the complexity is actually $\bigo{n^6}$. Our automated system correctly gives the $\bigo{n^6}$ analysis.
As another example, \citet[\S 5.4]{kuhlmann-etal-2011-dynamic} discovered a surprising property of their tabular dependency parsing algorithm.  Their initial naive analysis gave time complexity of $\bigo{n^5}$ and space complexity of $\bigo{n^3}$, but they were able to tighten this analysis to a time complexity of $\bigo{n^3}$ and space complexity of $\bigo{n^2}$, after noticing that the objects built by their transition system always had two equal variables (which could be proved inductively).  They then revised their program formulation to eliminate the redundant variable.  Our system automatically discovers the redundancy and the tighter analysis when run on the relevant program. 

\citet{eisner-blatz-2007} noted that Dyna programs can be automatically transformed in various ways that preserve their semantics but might affect their worst-case runtime.  To automatically search for sequences of transformations that \emph{improve} worst-case runtime, we would need to ascertain what that worst-case runtime is, e.g., by obtaining an asymptotic bound that is as tight as possible.\looseness=-1

This paper provides a theory for the automated analysis of Dyna programs as well as a working system for performing that analysis. 
Again, in the context of NLP, our automated analysis yields big-$\Bigo$ bounds on runtime and space complexity---the sort of bounds that were obtained manually by many NLP authors.\cutforspace{\footnote{\label{fn:mcallester-aficionados}%
This list includes the following citations: \citet{gildea-2011-grammar,nederhof-satta-2011-prefix-probability,gilroy-etal-2017-parsing,melamed-2003-multitext,kuhlmann-2013-mildly,nederhof-sanchez-saez-2011-parsing,buchse-etal-2011-tree,lopez-2009-translation,eisner-blatz-2007, huang-sagae-2010-dynamic,kuhlmann-etal-2011-dynamic}.\looseness=-1}}
Concretely, this paper makes the following contributions:
\begin{enumerate*}[label=(\roman*)]
\item a notation for specifying and reasoning about types (\cref{sec:type-analysis}),
\item a novel algorithm for type inference for Dyna programs (\cref{sec:abstract-fc}),
\item a notation for specifying input size, 
\item a time and space complexity analyzer (\cref{sec:time-and-space-analysis}) that automates and refines the prevalent runtime and space analysis technique in the NLP literature, which is based on \citet{mcallester-2002-meta}.
\end{enumerate*}
Our paper also contains numerous examples from the NLP literature rewritten in Dyna and presents them in a unified manner in order to demonstrate that our proposed algorithm arrives at the same analysis as attested in the literature.\looseness=-1

\section{Dynamic Programming in Dyna}
\label{sec:dyna}

Dyna is a high-level, domain-specific language for concisely expressing dynamic programs as recurrence relations.  We briefly describe the language here, but refer the reader to \citet{eisner-goldlust-smith-2005} for a proper presentation.
We will start with simple or familiar computations as examples.

\begin{myexample}
Matrix--matrix multiplication plus the identity matrix.
\begin{dynaex}{}
a(I,K) += b(I,J) * c(J,K).
a(I,I) += 1.   @\label{line:infinite-identity-matrix}@
\end{dynaex}
\end{myexample}
\noindent This program defines a collection of derived \defn{items} \dinline{a(I,K)} in terms of other items
\dinline{b(I,J)} and \dinline{c(J,K)}, which may themselves be derived from similar expressions or inputs.  
For any $(\vI, \vK)$ pair, the program defines the \defn{value}
\dinline{a(I,K)} to be $\big(\sum_{\vJ}$ \dinline{b(I,J)} $\cdot$ \dinline{c(J,K)}$\big) +1[\vI=\vK]$.  
The rules' use of \dinline{+=} means that the value is being defined as a summation.  The
first rule contributes one summand for each possible value of $\vJ$, since $\vJ$ appears only on the right-hand side of that rule.  The second rule contributes the additional summand $1$ in the case where $\vI=\vK$.

Dyna rules may be recursive:
\begin{myexample}\label{ex:min-cost-to-go}
Minimum-cost path in a graph.
\begin{dynaex}{}
beta(J) min= stop(J).
beta(I) min= cost(I,J) + beta(J).
\end{dynaex}
\end{myexample}
\noindent This program defines \dinline{beta(I)} to be the cost of the minimum-cost path in a graph from the node \vI to any terminal node \vJ.  Here \dinline{stop(J)} must first be defined to be the cost of stopping at \vJ when \vJ is a terminal node (and should be left undefined otherwise).  Also, \dinline{cost(I,J)} must first be defined to be the cost of the directed edge from \vI to \vJ (and should be left undefined if there is no such edge).

\begin{myexample}\label{ex:cky}
Weighted context-free parsing with CKY \citep{cocke-schwartz1970,younger67parsing,kasami65cky}, or, more precisely, the inside algorithm \citep{baker79trainable,jelinek1985markov}:\cutforspace{\footnote{If the reader is not familiar with context-free parsing, we recommend \citet[chapters 12--13]{jurafsky-martin-book}.}}\timv{A reviewer pointed out that the grammar is in canonical two form \citep{harrison1978introduction}. They also pointed out that the algorithm is technically not "CKY" because it doesn't require that the grammar be in CNF.}
\begin{dynaex}{}
beta(X,I,K) += gamma(X,Y,Z) * beta(Y,I,J) * beta(Z,J,K).
beta(X,I,K) += gamma(X,Y) * beta(Y,I,K).
beta(X,I,K) += gamma(X,W) * word(W,I,K).
goal += beta(s,0,N) * len(N).
\end{dynaex}
\end{myexample}
\noindent This program expects the 
\dinline{gamma},
\dinline{word},
and \dinline{len} items to encode specific things.
The \dinline{gamma} items should be the weights of the corresponding productions in the context-free grammar (e.g., \dinline{gamma(np,det,n)}$\,=\,0.8$ encodes the production $\dd{np} \xrightarrow{\tiny 0.8} \dd{det}\, \dd{n}$).
The item \dinline{word(W,I,K)} should be 1 if $\vW$ is the word at position $\vK$ of the input sentence and $\vI=\vK-1$, and should be left undefined otherwise.
Lastly, \dinline{len(N)} should be 1 if the sentence has length \vN.
Given such inputs, the item \dinline{beta(X,I,K)}
will represents the total weight of all grammatical derivations of the nonterminal $\vX$
over the substring spanning positions $(\vI, \vK]$.
Additionally, \dinline{goal} will be the total weight of all grammatical derivations of the entire input sentence.

\newcommand{\Groundings}[0]{\Gamma}

More formally, a Dyna \defn{program} $\Prog$ serves to define \defn{values} for \defn{items}.  Each item is named by an element of the ``Herbrand universe'' $\theHerb$---that is, a nested named tuple such as \dinline{f(g(z,h(3)))}, which is a 1-tuple of type \dinline{f} whose single element is a 2-tuple of type \dinline{g}.  The elements of $\theHerb$ are known as \defn{ground terms} because they contain no variables.\cutforspace{\footnote{Terms are a common construct in logic programming languages, such as Prolog \citep{prolog}.}}
The program $\Prog$ itself is an unordered collection of \defn{rules},\cutforspace{\footnote{Note that the collection may contain duplicates.}}
typically of the form $\dhead \;\dopluseq\; \db_1 \,\dotimes \myDots \dotimes\, \db_K$. 
We call $\dhead$ the head, and $\db_1, \myDots, \db_K$ the body of the rule.
Each $\db_k$ in the body is called a \defn{subgoal}.  

The operations $\doplus$ and $\dotimes$ manipulate values from a set $\Values$ and may be any pair of operations that form a \defn{semiring}\timv{For lifted inference in the semiring, we will also need a ``sum copies'' shortcut: $K \!\cdot\! v \defeq \bigoplus_{k=1}^K v$ for all $v \in \Values$ and $K \in \mathbb{N} \cup \Set{\infty}$. There is a default implementation for $K < \infty$.
}
\citep{goodman-1999-semiring,huang-2009-dynamic}.  
Let $\semizero\,$ and $\,\semione\,$ denote the semiring's additive and multiplicative identity elements, respectively.  If an item's value is left undefined or has no $\oplus$-summands, then this is equivalent to saying it has value $\semizero$, which means it cannot affect the values of any other items; in effect, it does not exist.
Common semirings include
total weight $\tuple{\mathbb{R}_{\ge 0}, \dplus, \dtimes}$, 
boolean $\tuple{\Set{ \dinline{false}, \dinline{true} }, \vee, \wedge}$,
minimum cost $\tuple{\mathbb{R} \cup \Set{ \infty }, {\color{aggblue}\texttt{min}}, \dplus}$, 
and Viterbi $\tuple{\mathbb{R}_{\ge 0}, {\color{aggblue}\texttt{max}}, \dtimes}$.\cutforspace{\footnote{Semirings provide an elegant abstraction for computing many common quantities in NLP modes,
such as expected values, gradients, and $K$-best \cite{goodman-1999-semiring,huang-2009-dynamic,eisner-goldlust-smith-2005,huang-chiang-2005-better}.}}
The distributive property of the semiring ensures an important interpretation of the value of each item as the sum over all proofs of that item, where each proof's value is the product of the values of its axioms.
In this paper, we assume that all rules in the program use the same semiring.\footnote{We leave the extensions of Dyna 2 \cite{dyna2}, which relaxes this restriction, to future work.}$^,$\footnote{We make one exception: boolean rules may be added to the program in addition to rules in a different semiring.  When converting from a boolean into the other semiring, 
we map \dinline{false} and \dinline{true} to $\semizero$ and $\semione$.  Going from the semiring into the space of booleans requires the \dd{?} operator (described later).}%

Rules do not have to name specific items.  The terms in a rule may include capitalized variables such as $\vX$ that are universally quantified over $\theHerb$.  This allows them to pattern-match against many items.
Let $\vX_1, \ldots, \vX_N$ denote the distinct variables in a rule $r$. Then a \defn{grounding} of $r$ is a variable-free rule $r[\vX_1 \mapsto v_1, \ldots, \vX_N \mapsto v_N]$ (that is, the result of substituting $v_i$ for all copies of $\vX_i$ in $r$, for each $i$) for some $v_1,\ldots,v_n\in\theHerb$. $\Groundings(r)$ denotes the set of all groundings of $r$.  Including a nonground rule in $\Prog$ is equivalent to including all of its groundings.

Dyna allows logical \defn{side conditions} on a rule, 
e.g., \dinline{goal += f(X) for X < 9}.
This is syntactic sugar for \dd{goal += f(\vX) * lessthan(\vX,9)}, 
where the value of each \dd{lessthan($a$,$b$)} item is $\semione$ if $a\! <\! b$ or $\semizero$ otherwise.   The logical expression $\dd{?}x$ has value $\semione$ if $x$ has any proofs, and $\semizero$ otherwise.\label{def:question-mark-operator}
The \dinline{lessthan} is an example of a \defn{built-in constraint} or \defn{builtin} for short.
Other builtins include \dinline{Z is X+Y}, which desugars to \dinline{plus(X,Y,Z)}.
Builtins are items that do not need to be defined in the program as they are part of Dyna's standard library.

\defn{Input data} $\Data$ is represented as a collection of \defn{axioms}---i.e., rules of the form $(\dhead \,\dopluseq\, v)$ where $v \!\in\! \Values$.
Adding these special rules to $\Prog$ results in a new program $\Prog'$.

\newcommand{\Step}[0]{\mathbf{T}}
\newcommand{\vv}[0]{\nu}

\paragraph{Semantics.}
A \defn{valuation} $\vv: \theHerb \to \Values$ is a mapping from ground terms to their values in the semiring.  Typically, $\vv$ assigns $\nonsemizero$ values to only a finite subset of $\theHerb$.
The \defn{step operator} $\Step_\Prog$ for a program $\Prog$ is a mapping from a valuation $\vv$ to a revised valuation $\vv'$.  
Essentially, $\Step_\Prog$ computes a new valuation by using the rules of $\Prog$.  
More formally, for any $\dhead \in \theHerb$,
\begin{equation}
\Big[ \Step_\Prog(\vv) \Big](\dhead) \defeq \smashoperator{\bigoplus_{\substack{ 
r \in \Prog, \\
(\dhead \,\dopluseq\, \db_1 \dotimes \cdots \dotimes \db_K) \in \Groundings(r)
}}}
\vv(\db_1) \otimes \cdots \otimes \vv(\db_K)
\label{eq:step}
\end{equation}

\noindent To run $\Prog$ on data $\Data$, we define $\Prog'$ to be the concatenation of $\Prog$ with $\Data$ and seek a fixed point $\vv^*$  of the $\Step_{\Prog'}$ operator.\footnote{\label{fn:axiom-interp}Since $\Data$ includes rules of the form $(\dhead \,\dopluseq\, v.)$ where $v \in \Values$ is a constant, we must interpret $\vv(v)$ in \cref{eq:step} as returning $v$.}\timv{re: \cref{fn:axiom-interp}. it's not limited to axioms, we should allow  \dinline{a += 1.2 * b}.}
In other words, $\vv^*$ is a solution to a certain system of nonlinear equations given by $\vv^* = \Step_{\Prog'}(\vv^*)$.  We then often write the value of an item $x$ as $\sem{x} \defeq \vv^*(x)$.  The system may admit multiple fixpoints or none at all.  
In practice, a suitable fixpoint $\vv^*$ is computed by some type of \defn{forward chaining}.\label{sec:ordinary-forward-chaining}  We will focus in this paper on the naive bottom-up algorithm, which simply initializes $\vv$ to the constant $\semizero$ function, and repeatedly updates $\vv$ to $\Step_{\Prog'}(\vv)$ until it converges to some $\vv^*$.\footnote{This may be improved to the \emph{semi}-naive bottom-up algorithm \citep{ullman88book,mcallester-2002-meta,eisner-goldlust-smith-2005}, but the difference is not important for our purposes.}
We will write $\Prog(\Data)$ as a shorthand for the final valuation $\vv^* \!=\! \Step_{\Prog'}(\Step_{\Prog'}(\cdots\Step_{\Prog'}(\semizero)\cdots))$ that is approached in this way.
The intermediate valuations essentially represent the evolution of the memo table (e.g., the parse chart in the case of parsing) as the dynamic programming algorithm runs.  In the boolean semiring, there is always a unique minimal fixpoint $\vv^*$, and for each item $x$, $\vv(x)$ reaches $\vv^*(x)$ in finite time.  However, there are programs where there is no finite time by which \emph{all} items have converged.

\paragraph{Utilities.}\label{sec:utils}
Lastly, we define some additional terminology and utility functions that will be useful later.  
Let $\support{\vv}$ denote the support of $\vv$, i.e., the set of items in $\vv$ with a \nonsemizero value.
Let $\head{r}$ and $\body{r}$ denote the head and body of the rule $r$.
Let $\vars{\cdot}$ denote the set of variables contained in a (possibly nonground) term, e.g., $\vars{\dd{f(g(\vX),4,\vX)}} \!\mapsto\! \Set{\vX}$.
Terms can be equated with other terms through 
structural equality constraints,
which are denoted by the equality operator, e.g., \dd{f(\vX) \!=\! f(3)}.  
Such constraints can be solved efficiently using a unification algorithm, 
$\Unify$ (e.g., \citet{robinson65unification,martelli1982efficient}), 
which returns a \defn{substitution} mapping $\vtheta$, 
which may equal $\emptyset$ if no solution exists.
For example, 
$\unify{\dd{f(\vY,4), f(g(\vX),\vZ)}}$ has the solution 
$\dd{\Set{\vY \mapsto g(\vX), \vZ \mapsto 4}}$.\footnote{We allow $\Unify$ to take a substitution mapping as an optional third argument, which enables algorithms to chain together the results of unification calls.}
When we apply the substitution to both sides (using the function $\Subst$), 
the terms become equal, $\subst{\dd{f(\vY,4)}}{\vtheta} \!=\! \subst{\dd{f(g(\vX),\vZ)}}{\vtheta}$.
On the other hand, \dd{f(\vX,g(\vX)) \!=\! f(3,g(4))} is unsatisfiable.  The notation $\vtheta_i$ returns the sub-map for the variables in a set $i$.  The notation $\vtheta(\vX)$ returns the value of the variable $\vX$ in the mapping (or \vX itself if no value exists in $\vtheta$).
Lastly, we use a function $\Fresh$, which renames all variables within a term, rule, or entire program to avoid variable name clashes.

\section{Type Analysis}
\label{sec:type-analysis}

\Jason{phrase more as "news you can use".  A section first about you, as the user, can specify the (upper-bound) types of your inputs.  Then, a section about how you get the (upper-bound) types of all items, via abstract FC---maybe starting by stating the answer that we \emph{want} to get on our CKY(?) example from the previous section.  The titles of the sections may be more practical: "How do I \ldots?"
Tim says that Chapter 6 is a better presentation, too.}

\Jason{are we trying to infer only user-understandable types here, so the only knots we can tie are knots that were declared as types by the user?  And if we want fancier type inference where the types are defined by tree automata with constraints, or Dyna programs, that's future work?}

\Timv{Need to assume range-restricted}

A \defn{type} is a subset of $\theHerb$---that is, a set of ground terms. In this paper, a type will be represented as a union of finitely many \defn{simple types}.  Each simple type is specified via a nonground term, whose variables may themselves be constrained to have particular types\jason{but types of what kind?  must they be parameters or other simple types (to support nesting)?  Is recursion allowed in a type definition---and is disjunction allowed---thus, can lists be simple types?  Where do user-supplied reasoning rules come in? In a conversation just now (8/13/24), Tim is taking the position that simple types can be anything that you'd be willing to emit as type analyzer output; but don't they also have to support closure properties, e.g., under intersection, so that when you combine the user's simple types, you still get a simple type?} or be constrained in other ways.  For example, \tinline{f(3,5)} is a ground term, but $\Set{\tinline{f(3,J)}}$ is an infinite set of terms specified using a variable, and
$\Set{\tinline{f(3,J)} \mid \tinline{ g(J), 3 < J} }$ or $\Set{ \tinline{f(I,J)} \mid \tinline{I < J} }$ is an infinite set whose specification involves a constraint.\jason{is \tinline{I < J} intended to imply that I and J are ints?} Each of these sets (simple types) can be conveniently specified by a single restricted Dyna rule (in the boolean semiring), as we will see below.

Reasoning about program behavior without actually running the program on concrete data is known in the programming languages community as \defn{abstract interpretation} \citep{cousot77abstract}.  Subsequent generation of compiled code can take advantage of any proof that an object manipulated by the code will have a particular type, both to represent the object efficiently as an instance of a class specialized to objects of that type, and to speed up tests such as pattern-matching.\jason{paper may need to say more about this}  If the program stores and iterates over a collection of distinct objects of a particular type, then knowing the cardinality of the type on the concrete data can yield bounds on the program's space and runtime requirements (\cref{sec:time-and-space-analysis}).\timv{Can mention dead rule/code elimination and lint checking;
we mention program optimization stuff in \cref{fn:compiler-optimizations}.}

This section explains what abstract interpretation looks like for Dyna programs.  Rather than computing the valuation $\Prog(\Data)$, we compute an upper bound on its support.  In other words, we infer a type that is guaranteed to contain all items that have a $\nonsemizero$ value.    Indeed, we will infer this item type without fully knowing the input $\Data$: the user will provide a type for the input items, using a notation we describe in this section, and from this guarantee, we will infer a type that is guaranteed to include all derived items as well.  

\newcommand{\BStep}[0]{\overline{\Step}}
\newcommand{\BData}[0]{\overline{\Data}}
\newcommand{\bx}[0]{\overline{x}}
\newcommand{\by}[0]{\overline{y}}
\newcommand{\bh}[0]{\overline{h}}
\newcommand{\ba}[0]{\overline{a}}
\newcommand{\bb}[0]{\overline{b}}
\newcommand{\bc}[0]{\overline{c}}
\newcommand{\bd}[0]{\overline{d}}

This section describes our type specification language and how it is used to reason about possible items.
Continuing with the CKY example, we provide the specification of the input data.  
However, as we are interested in the \emph{possibility} that an item may have a $\nonsemizero$ value, rather than finding its specific value, we first transform the program as illustrated below.

\begin{myexample} Recall, the program to analyze is\timv{we can save space by not repeating the code; reference \cref{ex:cky}; Note, however, that the earlier version did not include a params line.}\label{ex:type-cky}
\begin{dynaex}{}
params: word; len; gamma.
beta(X,I,K) += gamma(X,Y,Z) * beta(Y,I,J) * beta(Z,J,K).
beta(X,I,K) += gamma(X,Y) * beta(Y,I,K).
beta(X,I,K) += gamma(X,W) * word(W,I,K).
goal += beta(s,0,N) * len(N).
\end{dynaex}
\end{myexample}

\noindent Because no rules have been specified to define the \dinline{word}, \dinline{len}, and \dinline{gamma} items, one might think that these items receive no \dinline{+=} summands and thus have value \dinline{0}.  However, the \dinline{params} declaration says that these items will be defined by the program input.\footnote{\citet{dyna2} provide input to a Dyna program using a ``dynabase extension'' mechanism that can augment a program with arbitrary new rules, but in this paper we take the simpler approach of regarding a Dyna program as specifying a function of some semiring-weighted relations.}\timv{Should the discussion of params should move to section 2?\response{jason} maybe?  But it seems fairly integral.
\response{jason} also, maybe defer the discussion of dyna extensions that is currently in footnotes (but give a forward pointer).}

The CKY program $\Prog$ above can be automatically ``booleanized''
to yield the \defn{type program} $\BProg$ below.\label{def:boolean-relaxation}\footnote{The translation additionally maps any constant values to \dinline{false} or \dinline{true} as if we had applied the \dd{?} operator (\cref{def:question-mark-operator}).  Additionally, the \dd{?}-operator may be dropped because the type program is already boolean-valued.}
The ${\color{aggblue}\colondash}$ and ${\color{aggblue}\texttt{,}}$ symbols traditionally denote the $\dopluseq$ and $\dotimes$ operations in the Boolean semiring\cutforspace{ $\tuple{\Values,\oplus,\otimes,\semizero,\semione}\!=\!\tuple{\Set{\dinline{true},\dinline{false}}, \vee, \wedge, \dinline{false}, \dinline{true}}$}.

\begin{typeex}{}
params: word; len; gamma.
beta(X,I,K) :- gamma(X,Y,Z), beta(Y,I,J), beta(Z,J,K). @\label{line:cky-type:binary}\label{line:cky-type:start}@
beta(X,I,K) :- gamma(X,Y), beta(Y,I,K).    @\label{line:cky-type:unary}@
beta(X,I,K) :- gamma(X,W), word(W,I,K).    @\label{line:cky-type:preterm}@
goal :- beta(s,0,N), len(N).  @\label{line:cky-type:goal}@
\end{typeex}

\noindent The intention is that \tinline{beta(X,I,K)} will have value \texttt{true} in the type program for all \dinline{X,I,K} such that $\dinline{beta(X,I,K)}$ has a non-\semizero value in the CKY program.  Of course, each program needs input: if $\Prog$ is run on data $\Data$, then $\BProg$ must be run on a booleanized version $\BData$.  The valuation $\BProg(\BData)$ now effectively specifies an inferred type of items that \emph{might} be derived by the program.\footnote{For example,  $\BProg(\BData)(\tinline{beta(np,3,5)})\!=\!\texttt{true}$ if $\Prog(\Data)(\dinline{beta(np,3,5)}) \!\neq\! \semizero$---and possibly even if $\Prog(\Data)(\dinline{beta(np,3,5)}) \!=\! \semizero$, for example, if $\semizero$ can result from adding $\nonsemizero$ values under $\oplus$.  Such overestimates can arise because the boolean program does not track actual values.\looseness=-1
}
\timv{Here might be a good place to introduce notation for the inferred type
$\Shape^* \defeq \BProg(\BData)$.  We use $\Shape$ in section 4 as the representation of $\vv$ in abstract forward chaining.  On the other hand, $\Shape^*$ may be too jarring since section 2 had the valuation be a map $\nu$, and suddenly it is a program or at least typeset like a program.
}
\cutforspace{See \cref{prop:booleanization} in the appendix for a more formal statement.}

In order to abstract away from the specific input data $\Data$, users may \emph{directly} specify
the booleanized input data $\BData$ from the previous paragraph.  This $\BData$ may be any upper bound on $\support{\Data}$ and serves as a type declaration for the unknown input $\Data$.
The type defined by $\BProg(\BData)$
then continues to include all items 
in $\support{\Prog(\Data)}$, for any and every dataset $\Data$ for which this upper bound holds.  For example, $\BProg(\BData)$ still gives value \dinline{true} 
to \tinline{beta(np,3,5)} if \dinline{beta(np,3,5)} could possibly have a $\nonsemizero$ value.

An example of an input type declaration for CKY is the following boolean program $\BData$, in which each rule separately specifies a simple type:\!\!\!\!\!
\begin{typeex}{}
params: k; w; n. @\label{line:cky-type:input-params}@
word(W:w,I:n,K:n) :- I < K. @\label{line:cky-type:input-start}@
len(N:n).
gamma(X:k,Y:k,Z:k).
gamma(X:k,Y:k).
gamma(X:k,W:w). @\label{line:cky-type:input-end}@
\end{typeex}
\Crefrange{line:cky-type:input-start}{line:cky-type:input-end} specify 5 sets (simple types) whose union is a type that contains all input items.  The notation \tinline{word(W:w,I:n,K:n) :- I < K} in \cref{line:cky-type:input-start} is shorthand for
\tinline{word(W,I,K) :- w(W), n(I), n(K), I < K}, indicating that items in the set $\Set{\dinline{word(W,I,K)} \mid \dinline{w(W), n(I), n(K), I < K}}$ may have \nonsemizero values in the input data $\Data$.  
The predicates such as \tinline{n} could be defined by additional lines like
\tinline{n(I) :- 0 <= I <= 20},
which defines the possible sentence positions for sentences of length 20.
However, to abstract away from any particular sentence length, we omit such a definition, leaving \tinline{n} as a parameter.  Thus, the above program $\BData$ actually defines a \emph{parametric} type, just as \texttt{List[$\alpha$]} in many programming languages is a parametric type that depends on a concrete choice such as $\alpha=\texttt{string}$.
It depends on three other relations, which specify the allowed terminals, nonterminals, and sentence positions.  
The concrete input type would have the form $\BData(\tinline{k},\tinline{w},\tinline{n})$.  

Note that while $\BData$ is written as a union of simple types, $\BProg$ is not.  Our lifted forward chaining algorithm (\cref{sec:abstract-fc}) will derive the following rules that do specify simple types for the derived items of $\BProg(\BData(\tinline{k},\tinline{w},\tinline{n}))$:
\begin{typeex}{}
beta(X:k,I:n,K:n) :- I < K.   @\label{line:cky-type:derived-start}@
goal.                         @\label{line:cky-type:derived-end}@
\end{typeex}
Adding these rules to $\BData$ (which declares the type of \emph{input} items) yields the desired description of a type that covers all $\nonsemizero$-valued items. 
This augmented program uses the same parameters as $\BData$ and provides an upper bound on $\BProg(\BData(\tinline{k},\tinline{w},\tinline{n}))$.  The bound implies that we will only build \tinline{beta} items where \vX is a nonterminal (has type \tinline{k}), 
and the \vI and \vK variables are sentence positions (have type \tinline{n}) with \tinline{I < K}. Additionally, the bound says (reassuringly) that it may be possible to build a \dinline{goal} item.
If \dinline{goal} were impossible, we would have detected a mistake in the code.
Similarly, if the bound had included \tinline{beta(X:k,I:k,K:n)}, it would be a suspiciously loose upper bound,
since \tinline{I} should never be a nonterminal.\jason{relies on eyeballing; shouldn't we be able to write down that \tinline{I} should be a position?
\response{timv}
I would be happy to sneak in a type assertion language into the is paper if it fits.  
That might be a good addition for camera-ready.
\response{jason} Maybe punt a bit for now and say that a type assertion could be used to say that we expect a certain type to cover beta, and generate a warning and a runtime check otherwise.  It would probably look like
\tinline{:- beta(X!k,I!n,K!n)}.
}

In summary, our type analyzer is given the program $\Prog$ together with a parametric input type $\BData$, and computes a parametric item type such as \crefrange{line:cky-type:derived-start}{line:cky-type:derived-end} for the items derived by $\Prog$.
In effect, it runs the booleanized program $\BProg$ on the input upper bound $\BData$.  
However, as the item type is only required to be an upper bound, we will permit the analyzer to ``round up'' as it goes, as needed to keep the type simple.  

\Jason{Since concrete forward chaining can also propagate delayed constraints and variables (cf. R-exprs), I think our lifted version has only two special properties: rounding up and metalogical predicates.  The metalogical stuff only shows up if we allow nonground chart entries during non-lifted execution.  It concerns instantiation status (which is not visible within Dyna proper) and infects both $\BData$ and the inferred rules.  It should be possible for both to define (in effect) tree set automata to specify the instantiation shape of a variable.  Possibly this bleeds back into Dyna proper if we extend the language to allow sets.  I'll add text, but probably not indicate quite how fancy this can get (e.g., "list of anys"). 
\response{jason} We think we probably will start by requiring concrete forward chaining to be ground.  But late in the paper, we can say that the nice features of abstract forward chaining---nonground terms with delayed constraints on their variables---could be incorporated into the concrete algorithm as well, allowing us to run a wider range of programs.  An intermediate step is to say that the concrete algorithm does allow nonground terms but without any delayed constraints on the variables (other than equality constraints); perhaps that's enough to get speculation and magic?}
\Jason{More generally, the things that we are passing forward in forward chaining should be updates to relations (just adding tuples in the case of type analysis).  These will generally have a head and a body, where the body specifies constraints on the variables in the head.  In general they will be R-exprs, I guess.  In the nonground case, we might have something like {\tt assertion\_update(f(X,Y,Z) :- Y in int, (Y<N)) :- are\_alldifferent\_vars([Y,Z]), N in int}?  In other words, the head describes the constraint system that will be passed forward at runtime, and the body tells us something at analysis time about the shape of that runtime constraint system?}

\section{Abstract Forward Chaining}
\label{sec:abstract-fc}

We now develop our type inference algorithm, which generalizes ordinary forward chaining (\cref{sec:ordinary-forward-chaining}) in the boolean semiring.
In ordinary forward chaining, the valuation $\vv$ at each step is a finite collection of ground items.  Our generalization allows $\vv$ to be a finite collection of simple types at each step (and so is $\vv^*$).\jason{for final version, use $\bar{\vv}$ and $\bar{\vv}^*$ for boolean lifted inference!  That will make it smoother to go back to ordinary $\vv$ when we explain Dyna execution.} This generalization is necessary because we no longer have any finite upper bound on the input data with which to initialize forward chaining: the \tinline{n} parameter is an unknown relation, and the \dinline{<} constraint is an infinite relation.  
Instead, we specify parametric simple types that upper-bound the unknown and possibly infinite input data.  Forward chaining derives new simple types, which may also refer to relations like \tinline{n} and \dinline{<}.
Indeed, these relations may emerge in the result $\vv^*$, as we saw in \cref{line:cky-type:derived-start}.\looseness=-1

We provide the abstract forward chaining algorithm in \cref{sec:type-inference-algorithm}.
We address the challenges arising from constraint accumulation in \crefrange{sec:relaxation}{sec:tying-the-knot}.

\subsection{Type Inference Algorithm}
\label{sec:type-inference-algorithm}

\Jason{The discussion of abstract step operator is confusing (see discussion below).  I think we should break this presentation into two steps.  First discuss infinite relations and non-range-restricted rules, for example, \dinline{foo(I,K) :- I < K}, and give an algorithm that will work on infinite valuations specified as finite unions of sets (where those sets are of a certain kind, which we have to discuss).  Then talk about the fact that with params, we need a parametric valuation function, and point out that the same algorithm works, handling calls to the params as delayed constraints just as it did for the builtins.}

To generalize forward chaining in the boolean semiring, we generalize our representation of the valuation map $\vv$ from \cref{sec:ordinary-forward-chaining}.  For this semiring, $\vv$ would normally be represented as an explicit collection of ground items that have value \dinline{true}.  This representation will not work for our purposes because we need to represent unknown and infinite sets.
Hence, we will represent $\vv$ as $\bigcup_{s \in \Shape} s$, the union of finitely many simple types.  An example $\Shape$ for the CKY example from \cref{sec:type-analysis} is the collection of simple types on 
\crefrange{line:cky-type:input-start}{line:cky-type:derived-end}, which gives $\vv^*$.

In some sense, \crefrange{line:cky-type:start}{line:cky-type:input-end} already describe $\vv^*$, but indirectly and not as a collection of simple types.
Recall that a simple type is an expression of the form 
\tinline{h :- c_1,ldots,c_M} 
where each \tinline{c_m} is externally defined as either a built-in constraint or a parametric constraint, 
and may contain variables.\jason{in future there might be some other newly defined constraint, like an algebraic data type, which can't be flattened if it's recursive}
Note that \tinline{h} cannot appear in the body of any other simple type.
This form ensures that each simple type is self-contained and can be implemented as an object-oriented class.  

The pseudocode below implements the step operation \cref{eq:step} on a finite collection $\Shape$ of simple types:\footnote{Notice that the code uses a \emph{set} of rules.
In our implementation, this set replaces rules with their most general version, i.e., if there is a pair of rules $r,r'$ such that $r' \subseteq r$, we keep $r$.\Jason{explain how this is determined}
This ensures, among other things, that rules that are equivalent under variable renaming do not both appear in the set.  It can also help to achieve convergence.}

\begin{algorithmic}\small\label{def:abstract-step}
\LinesComment{Recall that the utility functions $\Unify$, $\Fresh$, and $\Subst$ were defined in \cref{sec:utils}.}

\LinesComment{Abstract step operator; Expands all rules of $\BProg'$ (i.e., $\BProg$ and $\BData$) against the current estimated type $\Shape$.  
Returns a set of simple types that denotes a type $\supseteq \Shape$.}
\Func{$\Inflate(\Shape)$}
    \State \Return $\bigcup_{r \in \BProg'} \Expand(\Shape, r)$
\EndFunc
\LinesComment{Expand the subgoals in $r$ against the simple types in $\Shape$. 
Returns a set of simple types.}
\Func{$\Expand(\Shape, r, k\texttt{=}1, \vtheta\texttt{=}\Set{}, \CC\texttt{=}\emptyset)$}
    \LinesComment{$k$: index of the current subgoal in $r$}
    \LinesComment{$\CC$: set of delayed constraints from subgoals $<k$}
    \LinesComment{$\vtheta$: substitution map}
    \State $(\est{\ddh} \ \colondash\ \est{\ddb}_1, \myDots, \est{\ddb}_K) \gets r$
    \LinesComment{Base case; no more subgoals}
    \If{$k = K\!+\!1$}
      \Return $\Set{ \subst{(\est{\ddh} \ \colondash\ \CC)}{\vtheta} }$
    \EndIf
    \LinesComment{Recurse on remaining subgoals}
    \State \Return $\displaystyle\smashoperator{\bigcup_{\hspace{25pt}\tuple{ \vtheta', \CC' } \in \LookupT(\Shape, \est{\ddb}_k, \vtheta)}}\hspace{-0pt} \Expand(\Shape, r, k\!+\!1, \vtheta', \CC' \!\cup \CC)$
\EndFunc
\LinesComment{Return the set of simple types in $\Shape$ who's heads unify with $\ddb$ (with the substitution mapping).}
\Func{$\LookupT(\Shape, \est{\ddb}, \vtheta\texttt{=}\Set{})$}
     \If{$\est{\ddb}$ is a param or builtin}
        \Return $\{ \tuple{\vtheta, \est{\ddb}} \}$
     \EndIf
     \State \Return $\{ \tuple{ \unify{\est{\ddb}, \est{\ddh}', \vtheta}, 
     \Set{ {\paramColor\ddc}'_1, \myDots, {\paramColor\ddc}'_M } }$
     \State \hspace{40pt}$\textbf{ for } (\est{\ddh}' \ \colondash\ {\paramColor\ddc}'_1, \myDots, {\paramColor\ddc}'_M) \in \fresh{\Shape} \}$
\EndFunc
\end{algorithmic}

Type inference is performed by fixpoint iteration on the $\Inflate$ operation (code given below).  This should be viewed as a ``lifted'' version of the forward chaining algorithm from \cref{sec:dyna}: it derives simple types from other simple types, rather than deriving ground items from other ground items.
The constraints that appear in these simple types are passed around during execution.  
Such constraints are known in the constraint logic programming literature as ``delayed constraints,'' in contrast to constraints like \dinline{3 < 2} that can be immediately evaluated and replaced with \dinline{true} or \dinline{false}.

However, the challenge that arises from the repeated application of $\Inflate$ is that constraints may accumulate.  They can accumulate indefinitely in the case of recursive rules since they effectively unroll an infinite chain of recursive calls, as will be illustrated in \cref{ex:recursive-constraints-accumulate-forever}.
To prevent divergence, we propose the following strategies:
constraint relaxation ($\Relax$; \cref{sec:relaxation}),
and constraint propagation ($\Propagate$; \cref{sec:rewriting-system}).
We discuss termination in \cref{sec:tying-the-knot}.
We give the overall algorithm below:

\begin{algorithmic}[1]\smaller
\Func{$\FixedPointT()$}
    \State $\Shape \gets \emptyset$   \label{alg:line:fp-init}
    \While{true}
        \State $\Shape' \gets \Relax(\Propagate(\Inflate(\Shape)))$
        \If{$\Shape' = \Shape$} \Return $\Shape$
        \EndIf
        \State $\Shape \gets \Shape'$
    \EndWhile
\EndFunc
\end{algorithmic}

\subsection{Relaxation}\label{sec:relaxation}

Consider the following input type $\BData$ for \cref{ex:min-cost-to-go}.

\label{ex:recursive-constraints-accumulate-forever}
\begin{typeex}{}
params: n.
beta(S) :- stop(S).
beta(S) :- edge(S,S'), beta(S').
stop(S) :- n(S).
edge(S,S') :- n(S), n(S').
\end{typeex}

\noindent Applying the $\Inflate$ operation once, which renames variables to avoid conflicts, we derive the following simple type for \dinline{beta} items (other types not shown):
\begin{typeex}{}
beta(S_1) :- n(S_1).   @\label{line:first-step}@
\end{typeex}
\noindent Applying $\Inflate$ a second time, we derive
\begin{typeex}{}
beta(S_1) :- n(S_1).             @\label{line:second-step-begin}@
beta(S_1) :- n(S_1), n(S_2).     @\label{line:second-step-end}@
\end{typeex}
\noindent Notice in the second rule that \tinline{n(S_2)} is lingering as a constraint from the recursive call.  In keeping with the (Prolog-like) semantics of boolean programs, the variable \tinline{S_2} is existentially quantified since it is a ``local'' variable that appears only in the rule body.  After applying $\Inflate$ $k$ times, we derive
\begin{typeex}{}
beta(S_1) :- n(S_1).
beta(S_1) :- n(S_1), n(S_2).
@$\vdots$@
beta(S_1) :- n(S_1), n(S_2), n(S_3), ldots, n(S_k).
\end{typeex}
This procedure is diverging by elaborating the type to describe paths of every length,
rather than converging to a fixpoint (i.e., a type that describes all vertices that can reach a terminal state on paths of \emph{any} length).
Our strategy for eliminating constraint accumulation from previous levels of recursion is to drop (i.e., relax) any constraint that refers to local variables:\timv{
Note that we could also tie the knot here by recognizing that
\tinline{b(S') :- n(S), n(S')} is a subset of \tinline{b(S) :- n(S)}.
}  

\begin{defin}[\mysf{relax}]
Let \tinline{h :- c_1,ldots,c_M} be a simple type. 
The operation \mysf{relax} drops each delayed constraint \tinline{c_m} that refers to a variable that does not appear in \tinline{h}.  In other words, we obtain a new simple type with the same head, but with the following subset of constraints $\Set{ \tinline{c_m} \mid \vars{\tinline{c_m}} \subseteq \vars{\tinline{h}}, m \in \Set{1, \myDots, M} }$.
To $\Relax$ a set of simple types, we simply map $\Relax$ over the set.
\end{defin}
\timv{Not sure I love the definition environment for this...}

Applying $\Relax$ to the second application of $\Inflate$ (\crefrange{line:second-step-begin}{line:second-step-end}),
we see that it would drop the second constraint of the second simple type because it depends on the non-head variable \tinline{S_2}, yielding:
\begin{typeex}{}
beta(S_1) :- n(S_1).
beta(S_1) :- n(S_1).
\end{typeex}
We may merge these rules because they are duplicates.
We now have the following type
\begin{typeex}{}
beta(S_1) :- n(S_1).
\end{typeex}
We now have a fixpoint:\timv{Maybe it should be called a ``relaxed fixpoint''
or a ``fixpoint of the relaxed step operator''?} this iteration's type (after relaxation) is equal to the previous iteration's type (\cref{line:first-step}).

Why is $\Relax$ valid? 
In general, dropping constraints from a simple type simply makes the type larger, raising the upper bound.  Dropping constraints that depend on non-head variables is simply a convenient choice.\timv{Should we move this paragraph elsewhere?}

We now consider an example where $\Relax$ leads to an overly loose type.
Recall CKY (\cref{ex:type-cky}).  The simple types from $\BData$ are added to $\Shape$ on the first $\Inflate$, such as

\begin{typeex}{}
gamma(X,W) :- k(X), w(W).
word(W,I,K) :- w(W), n(I), n(K), I < K.
\end{typeex}

\noindent On the next $\Inflate$, \cref{line:cky-type:preterm} from $\BProg$

\begin{typeex}{}
beta(X,I,K) :- gamma(X,W), word(W,I,K).
\end{typeex}

\noindent gives us an initial type for \dinline{beta}:

\begin{typeex}{}
beta(X,I,K) :- k(X), w(W), n(I), n(K), I < K.
\end{typeex}

\noindent After $\Relax$, we have

\begin{typeex}{}
beta(X,I,K) :- k(X), n(I), n(K), I < K.
\end{typeex}

\noindent However, on the next $\Inflate$, \cref{line:cky-type:binary} from $\BProg$
\begin{typeex}{}
beta(X,I,K) :- gamma(X,Y,Z), beta(Y,I,J), beta(Z,J,K).
\end{typeex}
will expand $\Shape$ to the following type:
\begin{typeex}{}
beta(X,I,K) :- k(X), k(Y), k(Z), 
  n(I), n(J), n(K), I < J, J < K.
\end{typeex}
$\Relax$ will drop all the constraints on local variables:\vspace{-\baselineskip}
\begin{typeex}{}
beta(X,I,K) :- k(X), n(I), n(K).
\end{typeex}
Unfortunately, relaxation resulted in the ordering information (\dd{<}) being lost.   In the next section, we will see how to combine \tinline{I < J} and \tinline{J < K} (before relaxing them) to deduce that \tinline{I < K}.  Since \tinline{I < K} only constrains head variables, it will survive the relaxation step.
This will allow the analyzer to deduce that \tinline{beta(X,I,K)} items \emph{all} have property \tinline{I < K} (rather than the first generation having \tinline{I < K}, the second having \tinline{I < J, J < K}, the third having e.g.\@ \tinline{I < J_1, J_1 < J_2, J_2 < J_3, J_3 < K}, etc.).

\subsection{Constraint Propagation}\label{sec:rewriting-system}

\Jason{the more rules the user writes, the better the bounds we can prove; so this is more like a proof assistant than something general that tries to find tight tree automaton types for implementation}

\defn{Constraint propagation rules} \citep{fruhwirth-1998}, are commonly used in constraint logic programming, such as ECLiPSe \citep{eclipse},
computer algebra systems, such as Mathematica \citep{mathematica}, 
and theorem provers, such as Z3 \citep{z3solver}.
In our setting, they provide a mechanism for users to supply domain-specific knowledge about type parameters and built-in constraints.  They are used to derive new constraints during inference, such as \tinline{I < K} above.

Below, we give motivating examples of constraint propagation rules and how they are useful for our running examples.\jason{maybe replace the ascii version of $\Leftarrow$ with the real math symbol, so it doesn't seem like we're writing down a system?}

\begin{itemize}[leftmargin=1em, topsep=1pt]
\item
In \cref{ex:cky}, the grammar start symbol \dd{s} is an element of the parametric nonterminal type \tinline{k}, and \tinline{0} is an element of the parametric sentence position type \tinline{n}:\cutforspace{\footnote{It is good practice to ensure that every constant in the program (such as \dinline{s} and \dinline{0} in the CKY program) is declared as a member or non-member of each primitive type.  This helps minimize the number of type constraints in the inferred type and, in some cases, it can facilitate discovering dead rules.}}\!\!
\begin{typeex}{}
k(s) <== true.  @\label{line:membership-k}@
n(0) <== true.  @\vspace{-.75\baselineskip}@
\end{typeex} 
\item 
To assert that the \tinline{k} and \tinline{n} types are disjoint:
\begin{typeex}{}
fail <== k(X), n(X).  @\label{line:disjoint-kw}@ @\vspace{-.5\baselineskip}@
\end{typeex}
Here \tinline{fail} is a special constraint that is never satisfied (a contradiction).
If \tinline{k(X)} and \tinline{n(X)} ever both appear in the body of a simple type, the simple type is empty and can be deleted.\timv{move later: Thus, during constraint propagation, if \tinline{fail} is ever derived, we may short-circuit propagation.}  Notice that \crefrange{line:membership-k}{line:disjoint-kw} are sufficient to derive a contradiction from either \tinline{n(s)} or \tinline{k(0)} (i.e., both are false).\looseness=-1
\item 
To assert that a type \tinline{d} is a subset of a type \tinline{c}:
\begin{typeex}{}
c(X) <== d(X).  @\vspace{-\baselineskip}@
\end{typeex} 
\item
To assert that \dd{<} is a strict partial order:
\begin{typeex}{}
(I < K) <== (I < J), (J < K). @\label{line:lessthan-transitive}@
fail <== (I < I).   @\vspace{-.5\baselineskip}@
\end{typeex} 
This addresses the problematic example in \cref{sec:relaxation}.
\item
To define a recursive type, such as a list\footnote{The bracket-pipe notation is a shorthand (borrowed from Prolog \citep{prolog}) for working with lists.  A list such as \dinline{[1,2,3]} desugars to \dinline{cons(1,cons(2,cons(3,nil)))}.  The notation \dinline{[X,Y|Rest]} would match as \dinline{X=1,Y=2,Rest=[3]}.} of nonterminals (i.e., type \tinline{k}):%
\begin{typeex}{}
ks([]) <== true.
k(X) <== ks([X|Xs]).@\label{line:listhead}@
ks(Xs) <== ks([X|Xs]).@\label{line:listtail}@
\end{typeex}
\end{itemize}

\timv{
With the exception of the recursive type rule, the constraint propagation rules above are useful in the CKY example to produce the derived types.  We provide another NLP application in \cref{ex:earley}, which makes use of the recursive type rule.
}

The $\Propagate$ method used in abstract forward chaining ($\FixedPointT$) runs a standard constraint propagation algorithm \citep{fruhwirth-1998}---which is just another forward chaining algorithm---to deduce all transitive consequences of the constraints in the body of a simple type.  The propagation rules should be chosen to be sound and reasonably complete, while still terminating in reasonable time.\timv{If the propagation rules are incomplete, we may infer a loose type, but we will never infer an incorrect type from sound propagators.}

\subsection{Removing Redundant Simple Types}\label{sec:redundant-simple-types}
We may remove a simple type $s$ from $\Shape$ if there exists a supertype $t \in \Shape$, that is, $s \subseteq t$.  Below is an (incomplete) method for testing such subtype relationships that leverages $\Propagate$.

\begin{algorithmic}

\Func{$\mysf{subtype}(s,t)$}
\LinesComment{$(s \subseteq t) \Leftrightarrow (s \cap t) = s$}
\State $r \gets \mysf{intersect}(s,t)$
\State $R \gets \Propagate(r)$
\State $S \gets \Propagate(s)$
\State \Return $R = S$  \Comment{equal up to renaming}
\EndFunc

\Func{$\mysf{intersect}(s,t)$}
\State $\vtheta \gets \unify{\head{s}, \head{t}}$
\State $s' \gets \subst{s}{\vtheta}$
\State $t' \gets \subst{t}{\vtheta}$
\State \Return $(\head{s'} \ \colondash\ \body{s'}, \body{t'})$
\EndFunc
\end{algorithmic}

Note that this test may fail to detect the subset relation if the rules for $\Propagate$ are incomplete.  (For example, to ensure termination, we may omit the sound rule \tinline{ks([X|Xs]) <== k(X), ks(Xs)} that reverses \crefrange{line:listhead}{line:listtail}.)  Future systems might use more powerful theorem provers, such as Z3 \citep{z3solver}, to test the subset relation.\jason{Try for final version, including implementation: We ourselves could easily replace the $r=s$ test to instead check whether each constraint in $r$ can be proved from $s$ by backward chaining on rules like the one cited, showing that $s$ is actually a subtype of $r$ and not just vice-versa, even if $r$ has constraints that don't appear in $s$.}

\timv{
The recipe for dead rule elimination:
For each $r \in \BProg$, if $\Expand(r) = \emptyset$, then we can safely remove $r$ from the $\Prog$.
}\timv{
There is a dual to dead rule elimination, which see if the rule is reachable from any declared queries.  This requires some similar but different machinery to check.  (An alternative strategy is to apply magic on $\BProg$ and then to do the test above.)}

\subsection{Termination}
\label{sec:tying-the-knot}\label{sec:truncation}

To prevent abstract forward chaining from building infinitely many simple types that correspond to ever-larger terms, we can truncate them to an optional user-specified \emph{depth limit} $D$.
This depth limit will rewrite a simple type in the following way: any subterm that appears in the head at depth $D$ is replaced with a fresh, unconstrained variable.\jason{I tried to change this to "any non-variable subterm" so that variables in the head at depth $D$ can still be reused elsewhere.  But then we would not be able to cut off something like \tinline{f([1,2,3,4,5])} if it were expressed as  \tinline{f([1,2|Xs]) :- Xs=[3,4,5]}}
This strategy ensures that only a finite number of relaxed simple types can ever be built.  Thus, it is sufficient to ensure convergence.  However, truncation generally leads to large overestimates, as the free variable has type $\theHerb$.

\Timv{
Several papers \citep{talbot97set-based,fruhwirth91type,yardeni1991type} propose type systems for logic programs, but do not allow for variables to covary, meaning these systems cannot recognize the equalities. \citet{filardo-thesis} is an exception.
}

\Jason{Proposal: We can ensure termination by imposing a maximum depth on terms and a maximum number of distinct symbols introduced by constraint propagation rules.  If these bounds are finite, then there are only finitely many rules that can be generated for a given type program (using the symbols that appear in the program), and thus the step function can't continue to increase $\vv$ forever.}

\Timv{example below}

\Timv{
\paragraph{User-specified initialization.}
Notice that our fixpoint iteration procedure $\FixedPointT$ initializes with $\emptyset$ (\cref{alg:line:fp-init}) as in ordinary forward chaining (\cref{sec:ordinary-forward-chaining}).
It is possible for the user to effectively provide a type declaration for derived items in addition to input items by adding simple type declarations for them.
This idea is reminiscent of programming languages, such as Java or C++, which require users to specify types for intermediate values, not just the input data.
From there, abstract forward chaining runs as usual.  If the user's type was too small, it will find a correct larger type.  However, if the user's type was larger than necessary, then the types that our system will find will be larger than necessary as well. \response{jason} Can't we trust our inference and take the intersection? Isn't the idea that these type declarations are assertions that should be either proved statically or checked dynamically? \response{tim, continuing previous text} In both cases, the user's initialization will tend to speed up inference relative to an empty type as initialization.  In some cases, like the depth-based truncation example above, our system would diverge without the help of the user's initialization.  To see how initialization helps, we will trace through the example with the initial type.
}

\section{Automatic Space and Time Analysis}
\label{sec:time-and-space-analysis}

\newcommand{\sizetrue}[1]{\left| #1 \right|}
\newcommand{\sizebound}[1]{\left\lceil #1 \right\rceil}

Now we develop an automated method for analyzing the space and time complexity of a Dyna program $\Prog$ given data $\Data$.
Given the types $\Shape$ associated with $\Prog$, the key is to find parametric upper bounds on their \emph{cardinality}. 
We will then use these to predict a parametric $\Bigo$-bound on the running time of forward chaining execution of $\Prog(\Data)$.

\subsection{Space Analysis}

We will use $\sizetrue{s}$ to denote the true cardinality of a set $s$ and $\sizebound{s} \geq \sizetrue{s}$ to denote an upper bound.

To bound the space complexity of $\Prog(\Data)$, it suffices to bound the total size of the item type $\vv = \bigcup_{s \in \Shape} s$ (\cref{sec:abstract-fc}).\jason{give authority for this: we only have to store the items plus indices that give only a constant factor blowup}  
Our high-level approach is $\sizebound{\vv} \defeq \sum_{s \in \Shape} \sizebound{s} \geq \sum_{s \in \Shape} \sizetrue{s} \geq \sizetrue{\vv}$,\footnote{If simple types in $\Shape$ overlap considerably, then the last inequality will be loose because summation double-counts the overlapping regions.  Removing subtypes (\cref{sec:redundant-simple-types}) addresses the most egregious case, but future work may consider transforming $\Shape$ into a \emph{partition} of simple types as follows. Replace any overlapping pair of simple types $s$ and $t$ with $s \cap t$, $s \smallsetminus t$, and $t \smallsetminus s$.  Repeat until no pairs $s$ and $t$ overlap.  Set difference $\smallsetminus$ can be implemented using negated versions of the elementary constraints.\looseness=-1}
 where the size $\sizetrue{s}$ of a simple type $s$ counts the number of satisfying assignments to its variables, and $\sizebound{s}$ is an upper bound on this.  

It is well-known that counting the variable assignments that satisfy a set of constraints can be computationally expensive. However, the more important issue in our case is that we do not even know how the parametric constraints are defined.
The remainder of this section considers how to construct a bound $\sizebound{s}$ from some user-supplied information about the parametric constraints.

\subsection{Input Size Specification}
\label{sec:size-specification}

\newcommand{\ii}[0]{{\color{vargreen}\mathbf{i}}}
\newcommand{\jj}[0]{{\color{vargreen}\mathbf{j}}}
\newcommand{\kk}[0]{{\color{vargreen}\mathbf{k}}}
\newcommand{\VV}[0]{{\color{vargreen}\mathcal{V}}}
\newcommand{\BB}[0]{{\color{vargreen}\mathcal{B}}}

\noindent Consider the simple type \tinline{gamma(X,W) :- k(X), w(W)}. 
A straightforward upper bound can be found: $\sizetrue{\tinline{gamma(X,W) :- k(X), w(W)}} \le \sizebound{\tinline{k(X)}} \cdot \sizebound{\tinline{w(W)}} = \Sk \cdot \Sw$ where $\Sk$ and $\Sw$ are user-specified size parameters.  (We will explain shortly how size parameters are specified.)
However, things become more complicated when there are multiple constraints on the same variable.  Consider the simple type
\begin{typeex}{}
beta(X,I,K) :- k(X), n(I), n(K), I < K.
\end{typeex}
It is clear that at most $\Sk\cdot\Sn\cdot\Sn$ assignments can satisfy the first 3 constraints.  Additionally, imposing the fourth constraint \tinline{I < K} can only reduce that number---even though the fourth constraint \emph{in isolation} would have infinitely many satisfying assignments ($\sizetrue{\tinline{I < K}}=\infty$).

We will describe how to derive an upper bound from the \emph{sequence} of constraints \tinline{c_1, ldots, c_M} in the body of the simple type $s$ (including any constraints added by propagation, as they may improve the bound).  Reordering the constraints does not affect the number of satisfying assignments; however, each of the $M!$ orders may give a different upper bound, so we will select the tightest of these.\timv{future: Additionally, if the user has specified a competing bound, we may consider it in the minimization.  \response{jason} Tim says he's thinking of a constraint on the number of rules in the grammar, or a constraint on derived items that the user knows how to prove but we don't; it's an assertion that we could check dynamically.  Also shows up with something like $\sizebound{\dinline{prime(X),even(X)}}=1$.}

The idea is to use \defn{conditional cardinalities}.  Let $V$ be any set of variables.  While \tinline{c} may be satisfied by many assignments to its variables, only some of these are compatible with a \emph{given} assignment to $V$.  In fact, suppose the user supplies a number $\sizebound{\tinline{c} \mid V}$ such that no assignment to $V$ can be extended into more than $\sizebound{\tinline{c} \mid V}$ assignments to $\vars{\tinline{c}}$.  The case $\sizebound{\tinline{c} \mid V}=1$ is of particular interest: this is necessarily true if $\vars{\tinline{c}}\subseteq V$ and asserts a functional dependency otherwise. 

For a simple type $s$ of the form $\tinline{h :- c_1, ldots, c_M}$, we can obtain an upper bound on the cardinality of $s$:
$\sizetrue{s} \leq \prod_{m=1}^M \sizebound{\tinline{c}_m \mid V_{<m}}$, where $V_{<m} = \bigcup_{i=1}^{m-1}  \vars{\tinline{c}_i}$.\footnote{Provided that all variables in the head \tinline{h} are constrained somewhere in the body.  If not, $\sizetrue{s}=\infty$.} 
More generally, an upper bound on the \emph{conditional} cardinality of $s$, given an assignment to some set of variables $V$, is $\prod_{m=1}^M \sizebound{\tinline{c}_m \mid V \cup V_{<m}}$.  Minimizing this over all orderings gives our final upper bound $\sizebound{s\mid V}$ (with $\sizebound{s} \defeq \sizebound{s \mid \emptyset}$).\timv{Give the improved memoized version, which uses suffix rather than prefix}

But where do the bounds $\sizebound{\tinline{c} \mid V}$ come from?  Consider the built-in constraint \dd{times}, which is an infinitely large relation on three integers.  Because \dd{times} has functional dependencies, we can take $\sizebound{\tinline{times(X,Y,Z)} \mid V}=1$ whenever $V$ contains at least two of $\{\tinline{X},\tinline{Y},\tinline{Z}\}$.  We allow a user to declare this conditional cardinality with the following notation:\vspace{-\baselineskip}
\begin{typeex}{}
|times(+X, +Y, Z)| <= 1. 
|times(X, +Y, +Z)| <= 1. 
|times(+X, Y, +Z)| <= 1.
\end{typeex}
For example, the third line says that given values for \tinline{X} and \tinline{Z}, the \tinline{times(X,Y,Z)} constraint will be satisfied by at most one value of \tinline{Y}.

Given all such declarations, we obtain $\sizebound{\tinline{c} \mid V}$ as the min of the upper bounds provided by conditional cardinality declarations that unify with \tinline{c} and whose variables marked with \tinline{+} are all in $V$.  This includes an implicit declaration that $\sizebound{\tinline{c} \mid \vars{\tinline{c}}} \leq 1$.  (Another implicit declaration is $\sizebound{\tinline{fail}} \leq 0$.)\looseness=-1

The same notation is useful for bounding input relations, such as the \tinline{word} relation in \cref{ex:type-cky}.
\begin{typeex}{}
|word(W, +I, K)| <= 1.
|word(W, I, +K)| <= 1. 
|word(W, I, K)| <= @\Sn@.
\end{typeex}
The first declaration says that there is at most one word \tinline{W}, paired with a single end position \tinline{K}, starting at each position \tinline{I}.  The second declaration is the mirror image of this. The third says that there are at most $\Sn$ words overall in the input sentence---where the symbol $\Sn$ is a \defn{size parameter}.  

We may also use size parameters to bound the cardinality of type parameters:\timv{we might want to move the simpler case earlier}

\begin{typeex}{}
|k(X)| <= @\Sk@.   |n(I)| <= @\Sn@.   |w(W)| <= @\Sw@.
\end{typeex}

\noindent which bounds the number of nonterminals by $\Sk$, sentence positions by $\Sn$ and terminals by $\Sw$.
Here $\Sk$, $\Sn$, and $\Sw$ are interpreted as symbols.
We will combine these symbols to form a \defn{parametric expression} that upper bounds the cardinality of the items in the program.  For example, as we saw earlier, $\sizebound{\tinline{beta(X,I,K)}} = \Sk \Sn^2$.  After adding up the sizes of simple types, our system can compute an asymptotic upper bound on the resulting expression, e.g., $\Sk \Sn^2 + \Sk \Sn + \Sk^2 \Sn = \bigo{\Sk \Sn \max(\Sn,\Sk)}$.

A final trick is to specify the cardinality or conditional cardinality of a composite constraint, which can be matched against a subsequence of $\tinline{c_1, ... c_M}$ when constructing an order-dependent upper bound.\jason{fill in details of the for the reader}  For example,
\tinline{|even(X),prime(X)| <= 1} says that there is only one even prime, while 
\tinline{|gamma(X,Y), k(Y)|} $\le \Sg$ 
says that no nonterminal symbol \tinline{X} can rewrite as more than $\Sg$ different nonterminal symbols \tinline{Y}.  (It may also be able to rewrite as many more \emph{terminal} symbols \tinline{Y}, but these would fail the \tinline{k(Y)} constraint.)

\subsection{Time Analysis}\label{sec:prefix-firings}

\newcommand{\suffixtime}[0]{\mysf{runtime}}
\newcommand{\prefixset}[0]{\mysf{prefix}}
\setlength{\marginparwidth}{1.6cm}

Suppose that the Dyna program is executed by\marginpar{\footnotesize\color{darkblue} Forward reference: See also \citet{eisner-2023-tacl}} the forward chaining algorithm of \citet{eisner-goldlust-smith-2005}.
This maintains two indexed data structures: the \defn{chart} and the \defn{agenda}. 
The chart represents the current estimate of $\vv$ as 
a data structure that maps from ground terms in $\theHerb$ (specifically, in the item type $\bigcup_{s\in\Shape} s$) to values in $\Values$.  
(Any item not in the chart is taken to have a value of $\semizero$.)
The agenda (or ``worklist'') is a queue of updates to be applied to the values of ground terms.  When an update to $\ddb$ is popped, its value is updated in the chart.  This change is then propagated to other items $\ddh$.  The updates to these $\ddh$ items are placed on the agenda and carried out only later.  Crucially, the affected $\ddh$ items are found by
matching $\ddb$ (the \defn{driver}) to any subgoal of any rule $\ddh\ \dopluseq \cdots$, and then querying the chart for items (\defn{passengers}) that match that rule's other subgoals.

Propagating an update from a given driver $\ddb$ through a rule $r$ requires us to find all the satisfying assignments to $\body{r}$ that are consistent with binding the driver subgoal to $\ddb$.  Although $r$ is not a type, a subgoal in its body, such as \dinline{gamma(X,Y)} does impose a constraint on \dinline{X} and \dinline{Y}---namely the constraint \tinline{gamma(X,Y)} from our type analysis (\cref{sec:type-analysis}), which is true only if \dinline{gamma(X,Y)} could have a value in the chart.

Suppose we have matched a driver $\ddb$ against a subgoal of $r$ and have already matched 0 or more other subgoals to passengers retrieved from the chart.  These matches have instantiated some variables $V$ in the rule.  For each subgoal match, we also know from our static type analysis (\cref{sec:abstract-fc}) that the variables involved in that match must have satisfied certain constraints $\CC$.  If we decide that \dinline{gamma(X,Y)} is the next subgoal that we will match against the chart, we will get at most $N \defeq \sizebound{\CC,\tinline{gamma(X,Y)} \mid V}$ answers.  Our methods from the previous section can compute this upper bound even though \tinline{gamma} is not a built-in or parametric constraint: simply refine the item type $\Shape$ by unifying the heads of all simple types $s \in \Shape$ with \tinline{gamma(X,Y)},\jason{we should remove redundant subtypes but the pseudocode currently doesn't} and then upper-bound the conditional cardinality of this refined type $\Shape'$.

Thanks to indexing of the chart, the runtime of getting and processing $N$ answers is only $\bigo{1+N}$ (with the $1$ representing the overhead of attempting the query even if there are no answers).\jason{shouldn't really use big-O notation here since we also care about small $N$.  Should say something like "at most a constant times $1+N$"}  For each answer, we must then proceed to match the remaining passengers in all possible ways.

The following pseudocode makes this precise.  Two subtleties are worth pointing out.  First, $\Shape$ is disjunctive.  The constraints $\CC$ that we accumulate from previously matched passengers depend on the particular simple types that cover those passengers, so we must iterate through all possibilities. Second, while the driver subgoal is chosen by the agenda,\jason{we can relax this if we're willing to use a non-agenda algorithm at runtime} the order in which to instantiate the other subgoals is up to us.  The pseudocode always chooses the next subgoal that will minimize the upper bound on runtime, given the simple types of the previously matched passengers.  In other words, instead of following the fixed subgoal order stated in rule $r$ \citep{eisner-goldlust-smith-2005}, we are using our static analysis to generate more flexible and efficient code.\looseness=-1

Let us assume that no item is popped more than once.  This can be arranged by topological sorting, if no item transitively depends on its own value\jason{our runtime bound is valid only for inputs that yield a cyclic hypergraph}
\citep{goodman-1999-semiring}.
Then our final runtime bound is $\sum_{r \in \Prog} \sum_{\ddb \in \body{r}} \sizebound{\,\est{\ddb}\,} \cdot \suffixtime(r, \Set{\ddb})$.\jason{this is wrong because constraints from $\ddb$ are never included}

\Timv{Change the pseudocode to use a set/list of simple types for the prefix - this enables us to keep track which subtypes are being used for each subgoal.}

\newcommand{\ddp}[0]{\mvar{p}}
\begin{algorithmic}\small\label{alg:runtime}
\LinesComment{
This recursive function (which should be memoized for efficiency) tells us the optimal ``suffix runtime'' for a rule
given that certain subgoals (the unordered set \prefixset) have already been grounded, with $\CC$ being the set of conjunctive constraints thereby accumulated.}
\Func{$\suffixtime(r,\prefixset\!=\!\emptyset,\CC\!=\!\emptyset,\vtheta\!=\!\Set{})$}

\LinesComment{Base case: all subgoals have been grounded out}
\If{$\body{r} = \prefixset$} \Return $1$  \EndIf

\State best $\gets \infty$
\State $V \gets \vars{\prefixset}$  \Comment{Variables bound by earlier subgoals}
\LinesComment{Optimize over the choice of next subgoal $\ddp$ from the set of remaining subgoals}
\For{$\ddp \in \body{r} \smallsetminus \prefixset$}
\State $t \gets 0$  \Comment{accumulate total suffix runtime (given choice $\ddp$)}
\For{$\tuple{\vtheta', \CC'} \in \LookupT(\Shape, \est{\ddp}, \vtheta)$}  \Comment{$\LookupT$ from \cref{sec:abstract-fc}}
\State $\CC'' \gets \Propagate(\CC \cup \CC')$
\If{\tinline{fail} in $\CC''$} \textbf{continue} \EndIf
\State $q \gets \sizebound{\CC', \est{\ddp} \mid V}$  \Comment{Maximum number of passengers}
\LinesComment{Recurse on the remaining subgoals}
\State $t$ \mysf{+=} $q \!\cdot\! \suffixtime(r, \prefixset \cup \Set{\est{b}}, \CC'', \vtheta')$
\EndFor
\State best \mysf{min=} $t$
\EndFor
\State \Return 1+best
\EndFunc
\end{algorithmic}

\newcommand{\pf}[0]{\mysf{prefix\_firings}}

\cutforspace{
\paragraph{Limitations.}
\emph{Constant-depth shapes}: Our approach can only reason about shapes where the depth of the term is bounded independently of the input data.  This means, for example, our system cannot infer a running time that is exponential in the length of an input list.  It would instead predict an infinite running time because type inference would be truncated (\cref{sec:truncation}).
\emph{Cycles:} 
Unfortunately, the running time of concrete weighted forward chaining is not proportional to the number of prefix-firings when cycles are present.  This is because forward chaining will run until a numerical fixed point, which can take an amount of time that is largely data (and semiring) dependent \citep{eisner-goldlust-smith-2005}.
We also note that the bound requires preprocessing the program \citep{mcallester-2002-meta} to avoid additional factors in the running time (e.g., a factor corresponding to the largest number of subgoals on the right-hand side of a rule).  This bound assumes that the implementation will spend a constant amount of time per rule grounding.\footnote{One algorithm due to \citet{goodman-1999-semiring} that matches the bound is bottom-up forward chaining on the unweighted program, followed by a second pass to compute the weights.
The forward-chaining algorithm of \citet{eisner-goldlust-smith-2005} uses a priority queue, which may cause updates to be repropagated more than a constant number of times if the priority function does not pop items in topological order.
}}

\vspace{-4pt}
\section{Conclusion}

\label{sec:related-work}\jason{future work: we've used Dyna notation, but Dyna itself supports updates: future work should consider runtime of propagating updates}

\Timv{
future work: 
better methods for tying the knot.  (infer tree automata)
better cardinality estimates by backtracking (like an approximate \#SAT solver).  
}

Many NLP algorithms can be expressed in Dyna; our proposed analyzer can automate much of the algorithmic analysis found in NLP papers, including deriving precise big-$\Bigo$ bounds on runtime and space complexity.
We have applied our system to a diverse set of algorithms in addition to those detailed in the paper.\footnote{These programs include 
several variants of dependency parsing \citep{huang-sagae-2010-dynamic,kuhlmann-etal-2011-dynamic,eisner-1996-three,johnson-2007-transforming,eisner-blatz-2007},
context-free parsing, including Earley's algorithm \citep{earley-1970},
Viterbi decoding \citep{viterbi67error},
hidden Markov models \citep{rabiner1989tutorial},
semi-Markov models \citep{sarawagi04semi},
edit distance \citep{Levenshtein},
Bar-Hillel's algorithm \citep{bar-hillel61grammars},
and many variants of shortest path in directed graphs.}
We have found type analysis to be invaluable in implementing and debugging structured prediction algorithms for NLP.
As we mentioned in \cref{sec:intro}, our system is capable of recovering some interesting cases that we found published in the NLP literature \citep{shi-etal-2017-fast,huang-sagae-2010-dynamic}.
In particular, we are able to recover  
\citet{shi-etal-2017-fast}'s improved analysis of \citet{huang-sagae-2010-dynamic} using our runtime bound inference, and \citet{kuhlmann-etal-2011-dynamic}'s time and space optimization of their exact arc-eager parsing algorithm.
We believe that this demonstrates a need for tools to facilitate the analysis and development of NLP algorithms.\looseness=-1

\bibliographystyle{acl_natbib}
\bibliography{main}

\end{document}